\def\heao1{{\it HEAO-1\/}}

\def\iras{{\it IRAS\/}}

\def\rosat{{\it ROSAT\/}}

\def\xmm{{\it XMM\/}}

\def\iras{IRAS~13224--3809}
\def\phl{PHL~1092}

\def\ltsima{$\; \buildrel < \over \sim \;$}
\def\simlt{\lower.5ex\hbox{\ltsima}}
\def\gtsima{$\; \buildrel > \over \sim \;$}
\def\simgt{\lower.5ex\hbox{\gtsima}}

\documentstyle[psfig]{mn}
 
\title[\rosat\ monitoring of extreme variability in \phl]
{\boldmath$ROSAT$ HRI monitoring of extreme X-ray variability in the narrow-line quasar \phl}

\author[W.N. Brandt et~al.]
{\parbox[]{6.5in}{W.N. Brandt,$^1$ Th. Boller,$^2$ A.C. Fabian$^3$ and M. Ruszkowski$^3$}\\
\\
$^1$ The Pennsylvania State University, 525 Davey Lab, University Park, PA 16802, USA\\
$^2$ Max-Planck-Institut f\"ur Extraterrestrische Physik, 85748 Garching, Germany\\
$^3$ Institute of Astronomy, Madingley Road, Cambridge CB3 0HA\\
}
 
\begin{document} 

\maketitle


\begin{abstract}
We report results from an 18-day \rosat\ HRI monitoring campaign on the 
ultrasoft Narrow-Line Seyfert~1 (NLS1) class quasar \phl. This luminous, 
radio-quiet quasar showed strong X-ray variability in a short \rosat\ PSPC 
observation, and \rosat\ HRI monitoring of the similar object
IRAS~13224--3809 revealed extreme variability on intermediate 
timescales. We wanted to determine whether remarkable X-ray variability 
persistently occurs in PHL~1092, and we also wanted to search for outstanding 
variability events that constrain emission processes. Given the large luminosity 
of \phl\ ($\sim 5\times 10^{45}$~erg~s$^{-1}$ in the HRI band), we 
detect extremely rapid and large-amplitude X-ray variability 
throughout our monitoring. The maximum observed variability amplitude is a 
factor of $\approx 14$, and in the most rapid variability event the HRI 
count rate increases by a factor of $\approx 3.8$ in a rest-frame time 
interval of $<3580$~s. The most rapid event has a rate change of luminosity 
of $>1.3\times 10^{42}$~erg~s$^{-2}$, making it the most extreme such event 
we are aware of from a radio-quiet quasar. Standard `radiative efficiency limit' 
arguments imply a radiative efficiency larger than can be achieved by accretion 
onto a Kerr black hole rotating at the maximum plausible rate, although we 
point out that such arguments depend upon the geometry of initial
radiation release. Relativistic motions of the X-ray source are probably causing 
the radiative efficiency limit to break down; such relativistic motions have 
also been inferred in the similar NLS1-class quasar PKS~0558--504. 
\end{abstract}



\section{Introduction}

\phl\ ($B=16.7$; $z=0.396$) is a luminous Narrow-Line Seyfert~1 (NLS1) 
class quasar that is one of the strongest optical Fe~{\sc ii} emitters known
(Bergeron \& Kunth 1980). Such objects have been generally found to 
have extreme X-ray spectral and variability properties, and it appears likely
that their exceptional X-ray/optical characteristics arise as the result 
of an extreme value of a primary physical parameter 
(see Brandt \& Boller 1998 for a recent discussion).
This primary parameter must ultimately originate from the immediate 
vicinity of the supermassive black hole since it can strongly influence
the energetically-important and rapidly-variable X-ray emission. 


\begin{figure*}
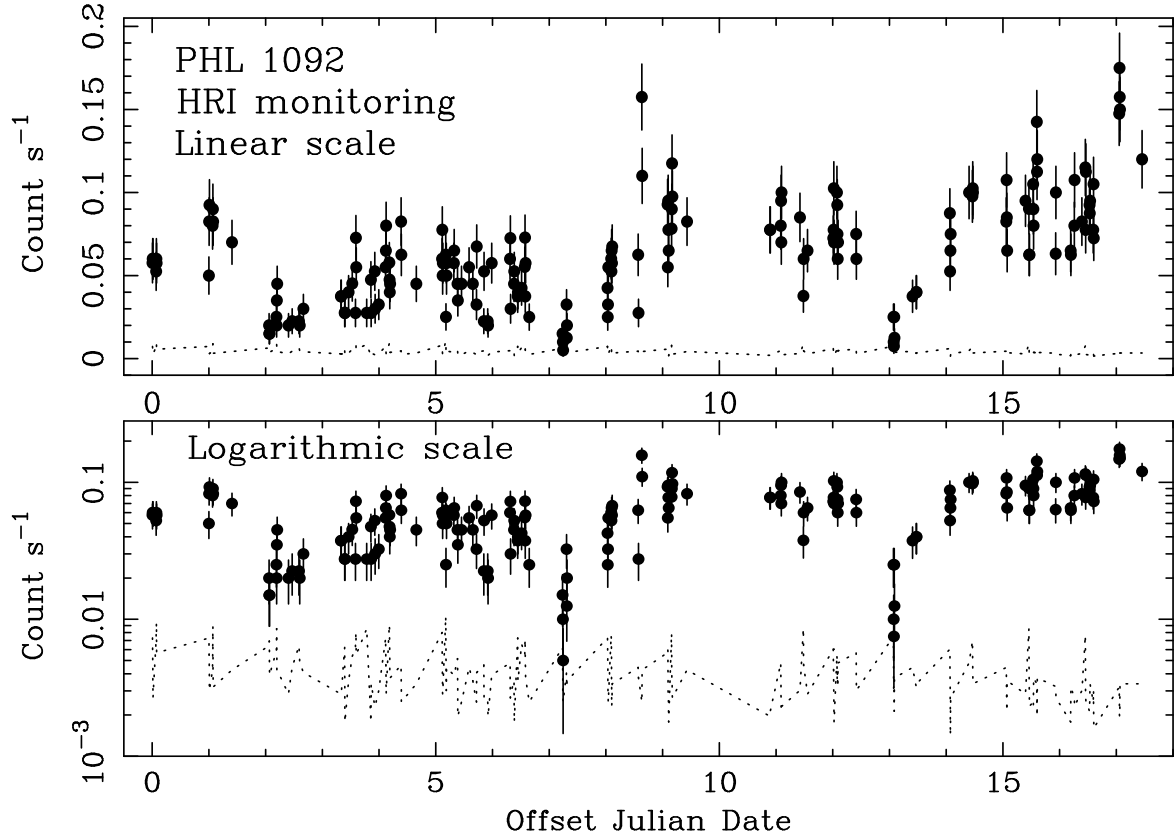

%
%
\hspace{0.20 cm} \psfig{figure=fig1a.ps,height=5.5cm,width=15.5cm,angle=-90,clip=} \\
\hspace{0.20 cm} \psfig{figure=fig1b.ps,height=5.5cm,width=15.5cm,angle=-90,clip=}
\caption{
\rosat\ HRI light curve for \phl\ obtained during the 18-day monitoring campaign 
between 1997~July~16 and 1997~August~2. The abscissa label gives the Julian date 
minus 2450645.12~days. The top panel shows the light curve with a linear
ordinate, and the bottom panel shows the light curve with a logarithmic ordinate. 
Each data point is plotted at the middle of the 400~s exposure interval from which it 
was obtained, and the sizes of the exposure intervals lie within the data points 
themselves. The dashed curve indicates the background counting 
rate within the source extraction circle as a function of time.
}
\end{figure*}


The \rosat\ PSPC spectrum of \phl\ is one of the softest ever seen from a 
quasar (Brandt 1995; Forster \& Halpern 1996, hereafter FH96; 
Lawrence et al. 1997). For example, a simple power-law fit to 
the PSPC spectrum gives a photon index of $\Gamma=4.2\pm0.5$. 
The poorly-sampled PSPC light curve showed 
remarkably rapid variability for such a luminous object. The count rate 
increased by a factor of $\approx 4$ during a 2-day period, and there
were weak indications for even more rapid variability. No strong 
spectral variability was apparent. FH96 used the `radiative efficiency limit' 
of Fabian (1979) to argue that the implied radiative 
efficiency was $\eta>0.13$, and they suggested that a Kerr black hole 
and/or anisotropic emission was implied. The other authors cited above 
interpreted the data in a somewhat more reserved manner (see below for
discussion), and they found $\eta\simgt 0.02$. In this case,
a Kerr black hole and/or anisotropic emission is not necessarily required. 

We performed an 18-day \rosat\ HRI monitoring campaign on \phl\ to 
further study its X-ray variability properties, and here we report the
results from our campaign. Our monitoring goals were  
(1) to determine whether extreme X-ray 
variability persistently occurs in this quasar 
and 
(2) to search for outstanding variability 
events even more extreme than those seen by the PSPC. 
As we discuss below, goal~2 is important since such variability events can
place constraints on emission processes and may elucidate the
origin of the extreme properties of ultrasoft NLS1 more generally. 
This work builds upon our HRI monitoring of the similar but lower-luminosity
ultrasoft NLS1 \iras\ (Boller et~al. 1997, hereafter BBFF). \phl\ is 
$\sim 40$ times more luminous than \iras. 

The Galactic column density towards \phl\ is 
$(3.6\pm 0.2)\times 10^{20}$~cm$^{-2}$ (Murphy et~al. 1996), and the
PSPC spectrum constrained the intrinsic column density to be
$<2.0\times 10^{20}$~cm$^{-2}$. 
We adopt $H_0=70$~km~s$^{-1}$~Mpc$^{-1}$ and $q_0={1\over 2}$, and we hence
derive a luminosity distance of 1840~Mpc. When we calculate luminosities 
below, we shall implicitly assume isotropic emission unless stated
otherwise. 


\section{ROSAT HRI monitoring results}

\subsection{Observations and spatial analysis}

\rosat\ HRI observations of \phl\ were performed between 
1997~July~16 (02:47:15~UT) and 1997~August~2 (13:53:58~UT). 
The total exposure time was 109.135~ks, and the source was centred 
on-axis. Up to five separate observations were 
obtained each day, and coverage was reasonably good except for
a period when \phl\ was not observable due the position of the Moon
(the largest gap in Figure~1). All data analysis has been
performed using {\sc ftools}.  

We have added together all the HRI observations to produce a master image. 
The centroid position of \phl\ in the HRI master image is 
$\rm \alpha_{2000}=01^h39^m56.1^s$, 
$\rm \delta_{2000}=06^{\circ}19^{\prime}24.7^{\prime\prime}$. 
This position is in good agreement with the precise
optical position given by Fanti et~al. (1977), and there is
no significant evidence for X-ray spatial extent after errors 
in attitude correction are taken into consideration (see Morse 1994). 
The source plus background photons at the HRI position of \phl\ 
were extracted using a circular source cell with a radius of 25~arcsec. 
The background was extracted using an annulus centred on 
\phl\ with an inner radius of 40~arcsec and an outer radius of 100~arcsec.

\subsection{Count rate variability}

In Figure~1 we show the full monitoring light curve for \phl. 
Strong X-ray variability is apparent throughout the light curve.
We have calculated the smallest and largest observed HRI count rates 
to determine the maximum variability amplitude. In such calculations,
we always average over at least 4 of the data points shown in 
Figure~1 to prevent statistical count rate fluctuations from 
inducing artificially large variability amplitudes. 
The smallest count rate is observed between days 7.2--7.4 in Figure~1, and 
the 6 data points during this span of time give a mean count rate of
$(1.13\pm 0.36)\times 10^{-2}$~count~s$^{-1}$. 
The largest count rate is observed between days 17.0--17.2, and  
the 4 data points during this span of time give a mean count rate of
$(1.57\pm 0.16)\times 10^{-1}$~count~s$^{-1}$. 
The most probable maximum variability amplitude is a factor 
of 13.9, and maximum variability amplitudes in the range
9.5--22.5 are most likely (Cauchy distributed; see section~2.3.4 of
Eadie et~al. 1971). 

In Figure~2 we show the most rapid observed variability event.
This event occurred around day 
8.6 in Figure~1. The HRI count rate rose from
$(3.46\pm 0.98)\times 10^{-2}$~count~s$^{-1}$
to 
$(1.31\pm 0.19)\times 10^{-1}$~count~s$^{-1}$, thereby
increasing by a factor of $\approx 3.8$ in $<5000$~s
($<3580$~s in the rest frame of \phl). We also 
detect 6 additional events where the HRI count rate 
exhibits a highly-significant increase or decrease by a 
factor $>2$ in $<1$~day. 


\begin{figure}
\psfig{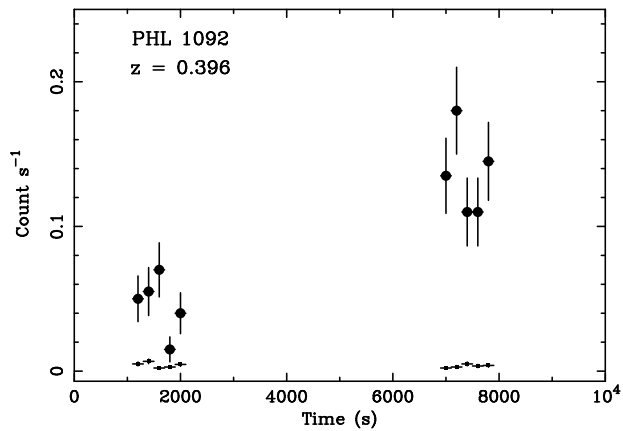}
\caption{
Light curve showing the most rapid observed variability during the
HRI monitoring campaign. The HRI count rate for \phl\ is shown by the 
large circular data points, and the background count rate in the source 
cell is shown by the small square data points. The bin size is 200~s. 
}
\end{figure}


\subsection{Luminosity variability}

We use the spectrum observed by the \rosat\ PSPC to determine
a conversion factor, $f$, between HRI count rate and  
luminosity. Unlike FH96, we do not simply extrapolate a steep
($\Gamma=4.2$) power-law model down to 0.1~keV to derive a 
0.1--2.0~keV luminosity. The extrapolation of a steep 
power-law spectrum to low energies where Galactic absorption is important 
can lead to unphysically large, or at least highly uncertain, 
luminosities. For such models most of the luminosity lies between 
0.1--0.2~keV where few counts are actually detected.\footnote{This 
is the cause of the much higher radiative efficiency derived by FH96 
as compared to Brandt (1995) and Lawrence et~al. (1997).} 

We have instead considered the luminosity from 0.2--2.0~keV in 
the observed frame since, for the Galactic column density relevant 
here, this quantity is much less dependent upon the 
details of the spectral model. 
We have considered three spectral models that give
statistically acceptable fits to the PSPC data: 
a power-law model (M1),
a power-law plus double blackbody model (M2) and
a power-law plus bremsstrahlung model (M3). 
Using {\sc xspec} (Arnaud 1996) and 
{\sc pimms} (Mukai 1997), we find
$f_{\rm M1}=6.0\times 10^{46}$~erg~count$^{-1}$, 
$f_{\rm M2}=4.8\times 10^{46}$~erg~count$^{-1}$ and 
$f_{\rm M3}=5.2\times 10^{46}$~erg~count$^{-1}$. 
In order to be reasonably conservative, we shall adopt $f_{\rm M2}$. 

We find that the 0.2--2.0~keV luminosity of \phl\ varies between
$(5.4\pm 1.7)\times 10^{44}$~erg~s$^{-1}$ and
$(7.5\pm 0.8)\times 10^{45}$~erg~s$^{-1}$ 
during our monitoring observation. 
If we assume that the optical continuum is not highly variable on 
timescales of a few weeks (cf. section~3.1 of FH96), the $\alpha_{\rm ox}$ 
value for \phl\ varies between $-1.8$ and $-1.4$ (we calculate 
$\alpha_{\rm ox}$ between 3000~\AA\ and 2~keV). For comparison, the 
$\alpha_{\rm ox}$ range for optically selected quasars is 
$\approx -1.5\pm 0.1$ (e.g. Laor et~al. 1997). 
For the variability event shown in Figure~2, we detect a change 
in HRI count rate of $(9.64\pm 2.14)\times 10^{-2}$~count~s$^{-1}$.
This corresponds to a luminosity change of 
$\Delta L=(4.6\pm 1.0)\times 10^{45}$~erg~s$^{-1}$ 
in a rest-frame time interval of $\Delta t<3580$~s. 


\section{Discussion}

\subsection{Radiative efficiency limit arguments}

The variability event shown in Figure~2 has
$\frac{\Delta L}{\Delta t}>1.3\times 10^{42}$~erg~s$^{-2}$, 
making it the most extreme such event we are aware of from a 
radio-quiet quasar (compare with Remillard et~al. 1991 and 
table~3 of FH96). This can be further quantified by employing 
the radiative efficiency ($\eta$) limit: 
$\eta>4.8\times 10^{-43}{\Delta L\over \Delta t}$
(Fabian 1979). Straightforward application of the limit gives 
an extremely high $\eta>0.62\pm 0.13$.
For comparison, optimal accretion onto a Kerr black hole rotating at the
maximum plausible rate gives only $\eta\approx 0.3$ (see Thorne 1974).

The extreme radiative efficiency derived above provides
motivation to critically examine the assumptions underlying
the radiative efficiency limit. The standard derivation 
assumes that the radiation release associated with an observed 
luminosity outburst occurs entirely at the centre of the emission region.
If the radiation release is more uniform, the 
rapid emission of photons from the outer few Thomson 
depths facing the observer can invalidate the standard derivation
(see Appendix~A). A rapid {\it rise} in the flux from a 
source, such as we have for \phl\ (and such as has been typically used 
in the literature), need not therefore involve the whole source and no 
limit applies. 

We can, however, recover the situation if either 
(a) we assume restrictions upon the manner of radiation release
or 
(b) we can place an upper limit upon the Thomson depth of the 
emission region. 
For case (a), we return to the original efficiency limit if all parts
of the region emitting the sharp flux increase are in causal contact
with each other (rather than with some remote centre of an idealized
sphere; see Appendix~A).
For case (b), we have from Appendix~A when $\tau_{\rm T}$ is large that 
$\Delta t = \vartheta {h\over c} = \vartheta {R\over \tau_{\rm T} c}$ 
where $\vartheta$ is a geometrical factor of order unity. 
Thus $\tau_{\rm T}=\vartheta {R\over c\Delta t}$. If we can place an 
upper limit on $R$ we also place an upper 
limit on $\tau_{\rm T}$. The emission dominating the 
\rosat\ band is from the soft X-ray excess, and this emission is
thought to be associated with the inner accretion disk. If we 
first make the assumption of blackbody emission we obtain
$R<4\times 10^{11} L_{44}^{1/2} T_6^{-2}$~cm where 
$L_{44}$ is the total blackbody luminosity
in units of $10^{44}$~erg~s$^{-1}$ and 
$T_6$ is the blackbody temperature in units of $10^6$~K.
We have $L_{44}\approx 50$, and $T_6$ is unlikely to be smaller 
than $\approx~0.35$ based on accretion disk theory and observations 
of the soft X-ray excess in similar objects (see equation 
3.20 of Peterson 1997; measured soft X-ray excess temperatures
are usually larger than this but will be strongly affected by
relativistic effects if the black hole is Kerr). We then find 
$R\simlt 2.3\times 10^{13}$~cm and $\tau_{\rm T}\simlt 0.5\vartheta$. 
While the derived $R$ constraint is physically plausible (of 
order a couple gravitational radii for a black hole mass of 
$\approx 10^8$~M$_\odot$), the low Thomson depth is inconsistent with 
the assumption that $\tau_{\rm T}$ is large. Thus when the
source size is determined by blackbody radiation, one 
cannot escape the efficiency limit by invoking large 
$\tau_{\rm T}$ as is done in Appendix~A. Additional opacity sources, 
such as free-free absorption, are expected in the case of blackbody 
or quasi-blackbody emission. These have the effect of
strengthening the above argument ($\tau_{\rm T}$ is replaced
by an effective $\tau_{\rm Tot}>\tau_{\rm T}$ that includes the 
additional opacity sources). 

Another possibility for case (b) is that the emission is due to
Comptonization (which is plausibly the fastest emission process 
in this band). In this case, a cloud of gas bathed in soft photons 
is impulsively heated. The soft photons must originate at energies 
below the HRI band for large variability to be seen. Large spectral 
variations will occur as the outgoing photons are successively Compton 
upscattered before escaping unless the electron temperature is large 
and $\tau_{\rm T}$ is small (see Guilbert, Fabian \& Ross 1982). 
The relative change in photon energy per 
scattering is $4kT/mc^2$, and therefore $kT\sim mc^2/4$ in order that many 
of the first burst of photons, which scatter at most once, appear in the HRI 
band. Such a high temperature requires that the Thomson depth of the whole
source is small. The equilibrium Compton scattering formulae given by
Zdziarski (1985; e.g. his equation 5; we correct his equation 6) require 
that $\tau_{\rm T}\sim 0.04$ if $kT/mc^2\sim 0.25$ in order to obtain
the steep mean photon index from the source of $\Gamma=4.2\pm 0.5$.
Again this upper limit on $\tau_{\rm T}$ allows a lower bound to
be placed upon $\eta$.

The precise value of the coefficient in the efficiency limit does depend
upon geometry and can be uncertain by a factor of a few. We have seen that
plausible constraints on radiation mechanisms make it unlikely that the much
larger uncertainty theoretically possible 
for a high $\tau_{\rm T}$ sphere is realised. We
note further that there are likely to be inefficiencies in the conversion of
the accretion energy into radiation, with some energy ending in kinetic or
magnetic form. We conclude that the rapid variation seen in Figure~2 requires
an efficiency of at least $\approx 0.10$ and more probably $\approx 0.60$.

\subsection{Relativistic X-ray boosting in ultrasoft NLS1?}

A straightforward application of the radiative
efficiency limit to our most rapid observed variability implies 
$\eta>0.62\pm 0.13$. This appears to be substantially larger than 
can be explained without allowing for relativistic effects, but 
moderate relativistic boosting can easily explain our data (see 
Guilbert, Fabian \& Rees 1983 and BBFF). The relativistic bulk 
motions required are of order $0.3c$ and might plausibly arise 
in the inner accretion disk. Relativistic X-ray boosting associated 
with a strong jet appears less likely given the spectral character 
of the soft X-ray emission and the fact that \phl\ is radio quiet. 

Evidence for relativistic X-ray variability enhancement 
has been found in two other NLS1-class objects to date: PKS~0558--504 and
\iras. PKS~0558--504 showed a rapid X-ray flare that implied relativistic 
motions (Remillard et~al. 1991), while \iras\ shows persistent, 
giant-amplitude, nonlinear variability that is most naturally explained 
via relativistic effects (BBFF). Our data on \phl\ now add further weight 
to the idea that unusually strong relativistic effects may be present in 
many ultrasoft NLS1. This hypothesis can be further examined by
searching for X-ray spectral changes during putative 
relativistic variability events. \xmm\ has the large collecting area 
and low-energy sensitivity needed for this work, and we hope to 
perform such observations.


\section*{Acknowledgments}
We thank the \rosat\ team for scheduling help. 
ACF acknowledges support from the Royal Society. 
MR acknowledges support from an External Research Studentship of
Trinity College, Cambridge; an ORS award; and the Stefan Batory Foundation.
This work has been supported by NASA grant NAG5-6023 and a NASA LTSA grant.


{}


\appendix

\section{Dependence of the radiative efficiency limit upon the 
geometry of radiation release}

The radiative efficiency limit was derived by arguing that a  
radiatively-inefficient source must have such a high particle density 
that a rapid release of radiation in the source appears to an observer 
to be slowed down by Thomson scattering. For a luminosity outburst of 
amplitude $\Delta L$ with a rise time of $\Delta t$, the 
radiative efficiency $\eta$ is defined by the equation

\begin{equation}
\Delta L\Delta t=\eta Mc^{2} 
\end{equation}

\noindent 
where $M$ is the mass involved in the outburst and $c$ is the speed of 
light. $M\approx m_{\rm p}nV$ where $m_{\rm p}$ is the mass of the proton, 
$n$ is the proton number density (also assumed to be the electron
number density), and $V$ is the volume of the emission region. 
The standard derivation of the radiative efficiency limit (Fabian 1979) 
assumes a uniform, spherical emission region with the release of 
radiation localized at its centre, and it also assumes that 
relativistic Doppler boosting and light bending are unimportant.
If the emission region has a significant Thomson depth 
($\tau_{\rm T}\ge 1$), $\Delta t$ must satisfy
 
\begin{equation}
\Delta t \ge (1+\tau_{\rm T})\frac{R}{c} ,\hspace*{1cm}\tau_{\rm T}=n\sigma_{\rm T}R
\end{equation}
 
\noindent
where $\sigma_{\rm T}$ is the Thomson cross section and $R$ is the radius of the region. 
The limit on $\eta$ arises due to the competition between the light crossing time 
and the photon diffusion time. By combining equations (A1) and (A2) one obtains
 
\begin{equation}
\eta \ge \frac{\Delta L}{\Delta t}f(\tau_{\rm T}), \hspace*{1cm} 
f(\tau_{\rm T})\propto \frac{(1+\tau_{\rm T})^{2}}{\tau_{\rm T}}. 
\end{equation}
 
\noindent     
%
$f(\tau_{\rm T})$ has the asymptotic behaviour
  
\begin{displaymath}
f(\tau_{\rm T})\propto \left\{ \begin{array}{ll}
        \tau_{\rm T}^{-1} & \mbox{for $\tau_{\rm T}\ll 1$ (light crossing time dominates)} \\
        \tau_{\rm T}      & \mbox{for $\tau_{\rm T}\gg 1$ (photon diffusion time dominates).} 
                        \end{array}
                \right.
\end{displaymath}
  
\noindent
This means that there exists a minimum value of $f(\tau_{\rm T})$ 
(4 when $\tau_{\rm T}=1$) and hence of $\eta$ for a given 
$\Delta L$ and $\Delta t$.

\begin{figure}
\centerline{\psfig{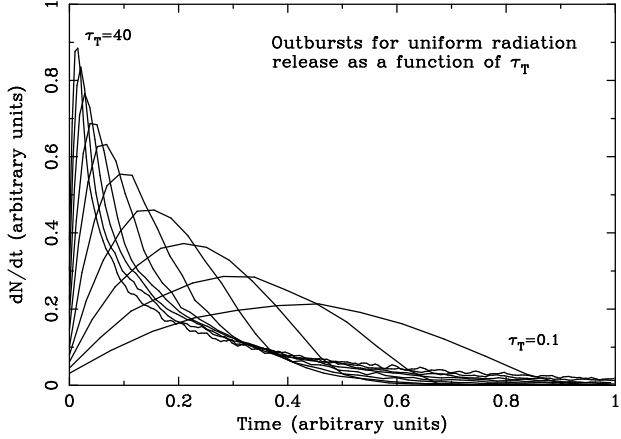}}
\caption{The photon flux detected at an 
observer $\frac{dN}{dt}$ versus time $t$ for 10 logarithmically spaced 
values of $\tau_{\rm T}$ from $\tau_{\rm T}=0.1$ to $\tau_{\rm T}=40$. Smaller 
Thomson depths correspond to the curves with lower maximum values.}
\end{figure}

In order to relax some of the assumptions entering the standard analytic
derivation, we simulated photon diffusion from `clouds' of moderate 
and high Thomson depths using Monte Carlo techniques. We first considered 
photon diffusion for the case described above with high Thomson depth 
and found our results to be consistent with the analytic results
published by Sunyaev \& Titarchuk (1980). We then considered a simple 
model of a uniform, spherical cloud but allowed radiation to be produced 
instantaneously throughout the cloud in a uniform 
manner. The results of the simulations for 
different Thomson depths are presented in Figure~A1. The curves show the 
time dependence of the observed photon flux $\frac{dN}{dt}\propto L$ for a 
wide range of $\tau_{\rm T}$. There is an important difference in the 
behaviour of these curves as compared to the case when the radiation 
release occurs entirely at the centre of the cloud: when $\tau_{\rm T}$ 
becomes large there is now no reduction in the rate of increase of 
$\frac{dN}{dt}$ at the start of the outburst. In other words, 
at the start of the outburst 
$\frac{d^2N}{dt^2}\propto \frac{dL}{dt}$ 
increases monotonically as $\tau_{\rm T}$ is increased 
(even for $\tau_{\rm T}\gg 1$). This result can be understood by realizing 
that the radiation observed near the start of the outburst comes mostly 
from the outer few Thomson depths of the cloud facing the observer. 
The relevant `cup-shaped' emission region at the start of the outburst is 
constrained by the surface of the cloud and the surface of equal 
photon arrival time. An upper limit to its volume for the diffusion dominated 
case is $V_{\rm cup}\le \pi R^{2}h$, where $h\approx\frac{R}{\tau_{\rm T}}$. 
Appropriate modification of equation (A3) leads to 

\begin{equation}
\eta \ge \frac{\Delta L}{\Delta t}\tilde{f}(\tau_{\rm T}), 
\end{equation}
where

\begin{displaymath}
\tilde{f}(\tau_{\rm T}) \propto \tau_{\rm T}^{-2} \begin{array}{l}
  \mbox{\hspace*{0.2cm} for $\tau_{\rm T}\gg 1$ (photon diffusion time dominates).} 
  \end{array}
\end{displaymath}

\noindent
The lower limit on the radiative efficiency in this case 
is a trivial $\eta \ge 0$.
 
The assumption that radiation is released instantaneously throughout the 
cloud is obviously an unrealistic one because it violates causality. 
To investigate the effect of causality, we considered a `trigger signal' 
propagating outwards from the centre of the cloud with speed $c$. 
Radiation is released in a volume element of the cloud only after the
trigger signal reaches it. Inclusion of this effect in the simulations 
does not change the general behaviour described above; at the start
of the outburst, $\frac{d^2N}{dt^2}$ increases monotonically
as $\tau_{\rm T}$ is increased. Again the only constraint on the 
radiative efficiency is the trivial $\eta \ge 0$.

A nonzero lower limit for $\eta$ can be derived if
one considers somewhat different definitions for $\Delta L$ and
$\Delta t$. For example, one could define the characteristic variation 
timescale $\Delta t$ as the time (measured from the observed start of the outburst) 
it takes for $\approx 90$ per cent of the outburst photons reach the observer. 
Similarly $\Delta L$ could be defined as 90 per cent of the energy liberated in 
the outburst divided by $\Delta t$. There is a nonzero lower limit for $\eta$ 
with the above definitions (this has been checked numerically). 
However, in order to reliably obtain an efficiency 
bound in this way one must be able to determine the overall profile of an 
outburst. In particular, one must be able to measure any decaying `tail' 
with reasonably high precision. This is usually not possible due to 
observational constraints and the complexity of active galaxy light 
curves (e.g. outbursts often overlap each other). 
 
In summary, the radiative efficiency limit is quite sensitive 
to the geometry of radiation release within the emission region. Without
constraints on this geometry or the Thomson depth of the emission region,
it is difficult to place meaningful constraints on $\eta$.



\end{document}